\documentclass[aps,prb,superscriptaddress,twocolumn,floatfix,showpacs,amsmath]{revtex4-1}
\usepackage{graphicx}
\graphicspath{ {./images/} }
\usepackage{floatrow}
\usepackage{latexsym}
\usepackage[makeroom]{cancel}
\usepackage{dcolumn}
\usepackage{epstopdf}
\usepackage{float}
\floatstyle{plaintop}
\restylefloat{table}

\usepackage{color}
\usepackage{amsmath,latexsym,psfrag}
\usepackage{array}
\usepackage{dcolumn}
\usepackage{bm}
\usepackage{epstopdf}
\usepackage{multirow}
\usepackage{soul}  
\usepackage[
colorlinks=true,
urlcolor=blue,
linkcolor=blue,
citecolor=blue]{hyperref}

\begin{document}
\begin{titlepage}
\title {First Principles Study of the Structural, Mechanical, Electronic,  and Lattice Dynamical Properties of the Half-Heusler Alloys ZrCoY (Y=Sb,Bi )}

\author{Lynet Allan}
\email {Corresponding author: allanlynet3@students.uonbi.ac.ke}
\affiliation{ Department of Physics, Faculty of Science and Technology, University of Nairobi, P.O.Box 30197-00100 Nairobi Kenya.}

\author{Winfred M. Mulwa}
\email{winfred.mulwa@egerton.ac.ke}
\affiliation{ Department of Physics, Faculty of Science, Egerton University, P.O Box 536-20115 Egerton Kenya} 

\author{Robinson J. Musembi}
\affiliation{ Department of Physics, Faculty of Science and Technology, University of Nairobi, P.O.Box 30197-00100 Nairobi Kenya.}


\author{ Bernard O. Aduda}
\affiliation{ Department of Physics, Faculty of Science and Technology, University of Nairobi, P.O.Box 30197-00100 Nairobi Kenya.}
\affiliation{ Department of Physics, Egerton University, P.O Box 536-20115 Egerton Kenya}

\date{\today}

 \begin{abstract}
First-principles calculations has led to significant discoveries in materials science. Half-Heusler (HH) alloys, which are potential thermoelectric materials have demonstrated significant improvements in thermoelectric performance owing to their thermal stability, mechanical strength, and moderate ZT. Using Density functional theory (DFT), the structutal, mechanical, electronic, and lattice dynamical  properties of cubic Half-Heusler Alloys ZrCoY (Y=Sb,Bi) have been investigated. The unknown exchange-correlation functional is approximated using the generalized gradient approximation (GGA) pseudopotential plane wave approach. The structural parameters, that is,  equilibrium lattice constant, elastic constant and its derivative are consistent with reported experimental and theoretical studies where available. Mechanical properties such as anisotropy factor A, shear modulus G, bulk modulus B, Young's modulus E, and Poison's ratio \textit{n} are calculated using the Voigt–Reuss–Hill average approach based on elastic constants. Debye temperature, as well as longitudinal and transversal velocities are predicted from elastic constants at GGA-PBE and GW levels of theory. The study of elastic constants showed that the compounds are mechanically stable, and the phonon dispersion study showed that they are dynamically stable as well. The ductility and anisotropic nature of the compounds were also confirmed by the elastic constants and mechanical properties.

\textbf{Key words}: {First Principles, Mechanical, Electronic, and Lattice dynamical Properties ,  Half-Heusler Alloys, ZrCoY(Y=Sb,Bi) }.

\end{abstract}
\maketitle
\end{titlepage}

\section{Introduction} \label{Sec :I}

Due to high energy demand, thermoelectric (TE) materials have recently attracted a lot of attention. TE compounds have been found to have the ability to convert heat to electricity in the first instance and electricity back to heat in the second instance. This concept is considerably adopted in cooling and power generation applications\cite{zhu2015high,schierning2015concepts,zeier2016engineering}. Excellent thermoelectric properties have been exhibited by different classes of compounds\cite{yu2012enhancement,niu2011enhanced}. Interestingly, semiconductor compounds have been identified as the most promising candidates for thermoelectric applications. Semiconductor elements are mainly found in group 3 to 5 of the periodic table. Binary alloys for example GaSb and AlSb exhibit semiconducting properties at room temperature\cite{strehlow1973compilation}. However, several ternary compounds present desirable semiconducting properties compared to binary and elementary semiconductors. Transition metal alloys are regarded as perfectly ordered as well as stable forms of the main class of ternary materials. These transition metal alloys are also known as Heusler alloys and they crystallize in cubic MgAgAs-type structure with space group number 216\cite{evers1997ternary}.  Heusler alloys are classified into two, that is, Full and Half-Heusler alloys. Full Heuslers (FH) have the atomic composition X$_2$YZ while half-Heuslers (HH) have atomic composition XYZ. In both FH and HH alloys, X and Y atoms represent transition metals, while Z stands for semiconductor atoms.
HH alloys have attracted researchers’ interests due to their wide practical applications in thermoelectric devices, spintronic and superconductors\cite{wei2017properties,wei2018thermoelectric,zeeshan2017ab}.  As a result, new HH compounds are always under study. Nonetheless, the  utility of HH alloys has been limited by mechanical and dynamical instability\cite{legrain2018materials}.  This work is mainly centred on investigating the structural, electronic, mechanical and dynamical stability of the HH alloys ZrCoSb/Bi, from first principles.

Recently, many researchers have focused on HH compounds comprising of eighteen valence electrons count by use of the semi-classical Boltzmann technique as well as the density functional theory technique. In particular, ZrCoSb has been subjected to both experimental and computational investigations\cite{xiao2018superconductivity,zhu2014electronic,sekimoto2006thermoelectric}.  Sekimoto and his team\cite{sekimoto2006thermoelectric} predicted the experimental lattice constant and band gaps of MCoSb (M=Ti,Zr,Hf)  using the experimental technique. Their finding showed that MCoSb is a semiconductor material, with promising thermoelectric properties. The findings were in good agreement with other experimental and computational data\cite{PhysRevB.59.8615,chauhan2019enhanced}. The most recent report on  HH alloys having a composition of MNiSn (M =Ti,Zr,Hf)\cite{sakurada2005effect,aliev1990narrow}, have lately emerged as  promising thermoelectric  device candidates.  Zr based HH alloys were found to have a higher figure of merit (ZT$>$1), compared to the other alloys in the study of  MNiSn and MCoSb (M=Ti,Zr,Hf) compounds. However, the stability of MNiSn and MCoSb HH alloys have not been explored. 

For temperature-sensitive applications (e.g thermoelectrics), understanding the mechanical and dynamical stability of HH compounds is essential. In fact, the ability of these materials to withstand intense, continuous temperature cycling and vibrational cracking improves their commercialization. The toughness and mechanical strength of a material are good indicators of crack resistance. One method for ensuring a structure's stability using theoretical calculations is to check its mechanical stability, which is a fundamental condition for thermodynamic and structural stability\cite{born1939thermodynamics}. Elastic constants play a significant role in determining mechanical stability and verifying the stability criterion and other parameters of a structure\cite{wu2017critical}. 

The quest for fundamental information on how HH materials behave under strain, has motivated us to explore the elastic properties such as elastic constants. These properties provide information that helps us to understand the thermal expansion, atomic bonding as well as structural, mechanical and dynamical stability of ZrCoY(Y=Sb,Bi) as potential thermoelectric materials.  The elastic constants, elastic moduli  and the lattice dynamics of the two HH compounds have also been calculated and analysed in this study.

The Phonon dispersion curves showed that the materials are dynamically stable and the band structures confirmed that the materials are narrow bandgap semiconductors. The following is a breakdown of the structure of this paper: Sec.\ref{sec : II} accounts for the technicalities that may be needed for future reproducibility in our computations. We offer detailed analysis of results and discussions in Sec.\ref{sec:III}. In Sec.\ref{sec : IIIA}, we discuss the structural properties of both ZrCoSb and ZrCoBi. The elastic constants and mechanical properties of the compounds under study are outlined in  Sec.\ref{sec :IIIB}, the electronic band structures and PDOS are discussed in Sec.\ref{sec :IIIC}, while Lattice dynamical (phonon dispersions) properties are discussed in Sec.\ref{sec :IIID}. The article's conclusions are outlined in Sec.\ref{sec :IV}.

\section{Computational details}\label{sec : II}

In this study, we have carried out simulations on the cubic HH ZrCoSb and ZrCoBi, both with a$=$b$=$c, $\alpha$=$\beta$=$\gamma$ = 90$^0$, within density functional theory (DFT)\cite{Kohn-65}. The open source Quantum ESPRESSO(QE)\cite{Giannozzi_2009} code was used. To accommodate the core electrons, the Perdew-Burke- Erzenhoff (PBE) flavor of the generalized gradient approximation (GGA) \cite{PBE-1996} functional was used. The electron-ion interactions were described by ultra-soft pseudo-potentials\cite{PAW-PPS}.   We relaxed the cell so that we can get the atomic positions at minimum energy, the atomic coordinates were relaxed until the forces were less than 0.01 eV/Å. After that structural optimization on lattice constant, cut-off energy and k-points followed. Experimental lattice parameters were considered as the starting point in the optimization of the lattice parameters. The data were fitted to the Birch-Murnaghan equation of state \cite{murnaghan1944compressibility}, with the required lattice constant obtained as the minimum value of the fitted curves. The optimal lattice constants are listed in table \ref{tab: table2}. Using the calculated lattice constants, the k-points were optimized by fixing the cutoff energy at 30Ry which is a small value that does not make the calculation computationally expensive during the test runs, the k-points were varied from a 2x2x2 grid to a dense mesh of 9x9x9.  The converged k-point meshes were obtained by considering the points of lowest energy from the graphs. Our optimum k-point grid was 5x5x5 for each of the two compounds. To obtain an optimal plane wave cutoff energies values, the lattice constants and k-point meshes were set at their optimized values then cut-off energy was varied from  20 Ry to 40 Ry, the optimum values were obtained at 30Ry. Hence, the wave functions for valence electrons were expanded through a plane-wave basis set within the energy cut-off of 30 Ry which limited the amount of plane waves we used. Running a self-consistent cycle iteratively solved the Kohn–Sham equations.  
The crystal structures were viewed using the X-window Crystalline and Molecular Structure Visualization (XCrysDen)\cite{kokalj2001xcrysden} program, the bondlenghs were studied and recorded in table \ref{tab: table 1}. A well-known problem of Kohn-Sham band structures obtained with local DFT functionals is the underestimation of the band gap. Using the  GW approximation, this problem can be resolved\cite{atambo2019electronic}. The GW approximation refers to  the expansion of the self-energy in terms of the single particle Green's function G and the screened Coulomb interaction W\cite{reining2018gw}. In most elemental or binary semiconductors, the conduction band is shifted upwards in an approximately rigid way\cite{reining2018gw}. In this study,  the GW calculation were performed  starting from the DFT (PBE) polarizability that calculates the screened interaction W, which in turn was used to determine the self-energy with the Green function G from DFT (PBE). Such calculations have been reported to be pretty reliable for semiconductor band gaps, which are underestimated by the Kohn-Sham gaps of local DFT calculations\cite{aoki2019insulating}. 

For the lattice dynamics, phonon dispersion were studied, six two-dimensional super cells with a total of 24 atoms were produced. The super cells were created using a 3-atom unit cell, with one atom each of Zr , Co and Bi/Sb for the two compounds under study. The Phonopy code\cite{togo2014phonopy} was used to calculate phonons for finite differences. 

The elastic constants were calculated using Quantum ESPRESSO's Thermo-PW post-processing code\cite{sekimoto2005thermoelectric}. Before doing all of the above computations, the material was subjected to variable cell relaxation to ensure that it was stress-free. The relax calculations were used to ensure the materials are not stressed. The Lagrangian theory of elasticity, in which a solid is considered to be an anisotropic and homogeneous elastic media\cite{voigt1910lehrbuch}, was used to compute the elastic properties within the DFT framework. Due to the cubic symmetry of the materials under consideration, there are three independent elastic constants.  Using the Voigt, Reuss, and Hill averaging approach\cite{voigt1910lehrbuch,reuss1929berechnung,hill1952elastic}, different elastic characteristics were computed using the elastic constants. In Voigt's approximation, the structure is assumed to have uniform strain, while  uniform stress is assumed in Reuss' approximation. 
Elastic moduli were described under the following averaging schemes:

The Bulk modulus B, which is the measure of resistance to compressibility was calculated using the expression in  Eq.\eqref{eqn1}.

\begin{equation}
    B = \frac{1}{3}  \big( C_{12} + 2C_{12} \big) \label{eqn1}
\end{equation}

For Voigt, Reuss, and Hill averages, the bulk modulus for a cubic structure is the same. The Shear modulus, being the deformation that occurs in a solid when a force is applied to one of parallel faces while the other face opposite the parallel face is held in place by opposing forces,  was calculated for the cubical symmetry in Voigt average using Eq.\eqref{eqn2}

\begin{equation}
    G_{V} =  \frac{C_{11} - C_{12} + 3C_{44} }{5} \label{eqn2} 
\end{equation}

The Reuss average was calculated using the expression presented in  Eq.\eqref{eqn3}.
\begin{equation}
G_{R} = \frac{ 5(C_{11} - C_{12} )  C_{44} }{4 C_{44} + 3 { (C_{11}} - C_{12} )} \label{eqn3}
\end{equation}

 The arithmetic mean of the Voigt and the Reuss average in  Eq.\eqref{eqn4} gave the Hill shear modulus
 \begin{equation}
     G_{H} =  \frac{G_{V}+ G_{R}}{2}   \label{eqn4}
\end{equation}
 Eq.\eqref{eqn5} and  Eq.\eqref{eqn6} were used to compute the Young's modulus and the Poisson's ratio 
\begin{equation}
    \gamma =  \frac{9BG}{3B+G}  \label{eqn5}
\end{equation}

\begin{equation}
      \eta  =  \frac{3B - 2G}{2(3B+G)}  \label{eqn6}
\end{equation}

By replacing G with G$_V$ and G$_R$ in  Eq.\eqref{eqn5} and  Eq.\eqref{eqn6}, the Voigt and Reuss averages of Young's modulus and Poisson's ratio were calculated.
Eq.\eqref{eqn7}, is based on Debye's assumption that the temperature of the greatest normal mode of vibration. The Debye's temprature was determined from the average sound velocity\cite{ozyar2015systematic}, as indicated in Eq.\eqref{eqn7}.

\begin{equation}
  \theta_{D} =  \frac{h}{k} \left[  \frac{ 3n }{4 \pi } { \left( \frac{pN _{A} }{M} \right)}  \right] ^{ \frac{1}{3} }  \mu _{m} \label{eqn7}
\end{equation}

where, $\textit{h}$ is the Planck's constant, $\textit{k}$ is the Boltzmann's constant, $\textit{N$_A$}$ is the Avogadro's number, $\textit{n}$ is the number of atoms per molecule or per formula unit, $\textit{M}$ is the molar mass, ${\rho}$ is the density of the unit cell, and ${\mu_m}$ is the average sound velocity.
Eq.\eqref{eqn1} further expresses the average sound velocity in terms of compressional (l) and shear (s) sound velocities \cite{anderson1963simplified}

\begin{equation}
    \mu _{m} =  \left(\frac{1}{3}\right) {\left[ {\frac{2}{ u^{3}_{s}}}+\frac {1}{u^{3}_{l}
    } \right] ^\frac{-1}{3} }   \label{eqn8}  
\end{equation}

Eq.\eqref{eqn9} and  Eq.\eqref{eqn10} respectively gives the expressions for $\mu{_s}$ and $\mu{_l}$

\begin{equation}
   \mu _{s}=  \sqrt{ \frac{G}{P} }  \label{eqn9} 
\end{equation}

\begin{equation}
   \mu _{l}=  \sqrt{ \frac{3B+4G}{3P} }  \label{eqn10} 
\end{equation}

The Shear anisotropy (A), a physical characteristic that is related to elastic properties was  calculated using  Eq.\eqref{eqn11} to determine the type of bonding in various crystallographic directions.

\begin{equation}
   A =   \frac{2C_{44}}{{ C_{11} - C_{12} } }  \label{eqn11}
\end{equation}

\section{Results and discussions}\label{sec:III}

\subsection{Structure of ZrCoSb and ZrCoBi}\label{sec : IIIA}

\begin{figure}[H]
    \centering
    \includegraphics[width=0.9\textwidth]{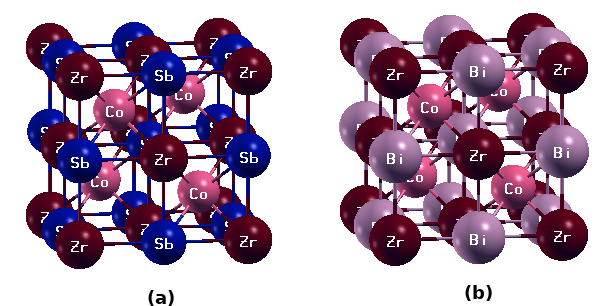}
    \caption{The crystal structures of Half Heusler alloys \textbf{(a)} ZrCoSb and \textbf{(b)} ZrCoBi.}
    \label{fig:Fig 1}
\end{figure}

A  cell relaxation calculation was used to achieve structural optimization, as well as K-points and cut-off energy. The  crystal structure's lattice parameters were estimated using data from prior studies of the ZrCoY compounds. By fitting the energy volume relationship to the Murnaghan equation of state\cite{murnaghan1944compressibility}, the volume that yields the minimum energy value are obtained.  Fig. \ref{fig:Fig 1} shows the crystal structure of the HH ZrCoY alloy, which consists of four interpenetrating face centered cubic (FCC) sub-lattices with the Zr, Co, and Sb/Bi atoms occupying the Wyckoff positions Zr(0, 0, 0), Co (1/4, 1/4, 1/4), and Sb/Bi (1/2, 1/2, 1/2), while the fourth sub-lattice located at (3/4, 3/4, 3/4) is vacant as shown in Fig. (\ref{fig:Fig 1}a)  and Fig. (\ref{fig:Fig 1}b). 

The bondlengths were then extracted and listed in table \ref{tab: table 1}. When compared to the bondlengths between Zr-Bi/Sb, the bondlengths between Co-Sb and Co-Sb/Bi are equivalent and shorter. Otherwise, the bondlengths are generally shorter, indicating that the compounds have a relatively high bulk modulus. 

\begin{table}[h]
   \begin{tabular}{llll} \hline 
         Bonds             & Zr-Co & Co-Sb/Bi & Zr-Sb/Bi \\
         Bond Lengths  (Å) & 2.64  & 2.64     & 3.05  \\ \hline 
     \caption{Calculated bond lengths for ZrCoY (Y=Sb,Bi)}
    \label{tab: table 1}
   \end{tabular}
\end{table}   

Our lattice constants for ZrCoSb  agree well with experimental values of 6.0676 Å  obtained by Sekimoto et al at room temperature\cite{sekimoto2005thermoelectric}.  However, the GW calculated lattice parameters gives an improved description of the compounds with reference to the experimental data. The values are listed in table \ref{tab:table 2}. The optimized and experimental lattice constant is in good agreement as shown in table \ref{tab: table2}. From table \ref{tab: table2} we note that the lattice constants of ZrCoSb compound is about 3\% less than that of ZrCoBi compound, an indication that the larger mass of Bi caused a little strain on the Co/Zr atoms.  Later, we will discuss how this difference affects the local bonding between the atoms on the resulting electronic structures since  the compounds are narrow-band-gap semiconductors, as predicted on the electronic band structures in Fig. \ref{fig:fig2} and Fig. \ref{fig:fig 3}.

\begin{table}[h]
\caption{Calculated lattice constants for ZrCoY (Y=Sb,Bi)}
\label{tab: table2}
\begin{tabular}{lllll}\hline
\multirow{3}{*}{Comp} & \multicolumn{4}{l}{Lattice Constant (Å)}                                         \\
                           & \multicolumn{2}{l}{This work} & \multirow{2}{*}{GGA} & \multirow{2}{*}{Expt} \\
                           & GGA         & GW          &                          &                       \\ \hline
ZrCoSb                     & 6.01            & 6.05        & 6.09\cite{joshi2019theoretical}                     & 6.06\cite{sekimoto2005thermoelectric}                  \\
ZrCoBi                     & 6.20            & 6.22        & -                        & -  \\ \hline                  
\end{tabular}
\end{table}

\subsection{Elastic constants and Mechanical Properties}\label{sec :IIIB}

Table \ref{tab:table 3} shows the elastic parameters that were calculated. The estimated elastic constant values indicate that both the compounds under investigation meet the Born-Huang stability criterion\cite{born1939thermodynamics}  given by ; 
C$_{11}$ $>$ 0, C$_{44}$ $>$ $0, $ C$_{11}$ -  C$_{12}$ $>$ 0, and  C$_{11}$ +  2C$ _{12}$ $>$ $0 $. As a result, the HH compounds ZrCoY (Y=Sb,Bi) are predicted to be mechanically stable. The  C$_{11}$ value for both ZrCoY compounds are higher than C$_{12}$ and  C$_{44}$, indicating that these compounds are difficult to compress along the X-axes. For the two compounds, ZrCoSb, and ZrCoBi, the Bulk modulus values computed using the Eq \eqref{eqn1} in the Thermo-PW code are 142.2 GPa and 142.1 GPa, respectively, which are  near the values calculated from Murnaghan's equation of state of 136.6 GPa and 136.1 GPa. The high bulk modulus of 142.2 Gpa indicates that ZrCoSb is incompressible, that is, it has high bond strength. The calculated shear moduli for ZrCoSb and ZrCoBi are 79.9 GPa and 71.3 GPa, respectively, which are good indicators for hardness. The GW computed elastic constants are however in fairly good agreement with the GGA-PBE calculations, an indication that the GW approximation does not cause major strain in bonding of the atoms in the systems. Unfortunately, there are no elastic constants as well as elastic moduli results for ZrCoBi, which makes comparison a challenge. The Young’s modulus of  202.6 GPa for ZrCoSb and 183.4 for ZrCoBi, confirms that ZrCoSb is stiffer than ZrCoBi. 
Bulk Modulus (B) and Shear modulus (G) are used to calculate the Pugh’s ratio (B/G), which is an indication of the ductility of materials. B/G $>$ 1.75, suggests ductility, whereas B/G $<$ 1.75 indicates brittleness in a material.  For ZrCoBi and ZrCoSb, the B/G ratios are 1.77 and 1.99, respectively, indicating that the materials are ductile. On the other hand, the poison's ratio (\textit{n}) $>$ 0.26, indicates the materials' ductility while \textit{n} $<$ 0.26 is an indicator for brittleness. The calculated values of  \textit{n} are 0.267 and 0.285, respectively, for ZrCoBi and ZrCoSb, confirming the ductility of these materials.  

\begin{table*}[t]
\caption{Calculated elastic constants, Bulk modulus (B), Shear modulus (G), Young’s modulus (E),Pugh's ratio (B/G), and
Poisson’s ratio \textit{(n)}}
\label{tab:table 3}
\begin{tabular}{llllllllll}
\hline\hline 
\textit{\textbf{Comp}}  & \textit{\textbf{C$_{11}$ (GPa)}} & \textit{\textbf{C$_{12}$ (GPa)}} & \textit{\textbf{C$_{44}$	(GPa)}} & \textit{\textbf{B$_{H}$ (GPa)}} & \textbf{G$ _{H}$(GPa)} & \textit{\textbf{B/G}} & \textit{\textbf{E $_{H}$(GPa)}} & \textit{\textbf{n}} & \textit{\textbf{Ref}} \\ \hline
\multirow{4}{*}{ZrCoSb} & 272.8  & 76.9  & 69.7  & 142.2  & 79.9 &  1.77  & 202.0   & 0.262   & This work, GGA-PBE    \\
                        & 275.5  & 79.9 & 69.7  & 144.9  & 79.9 &  1.81  & 202.6  & 0.267    & This work, GW         \\
                        & 273.0  & 78.1  & 70.7 & 140.8  & 79.3 &   1.77  & 202.3   & 0.25   & Ref\cite{joshi2019theoretical}, GGA-PBE      \\
                         & -    & -& -  & -   & -   &     & 201  & -   & Expt\cite{sekimoto2005thermoelectric}                  \\
\hline \multirow{2}{*}{ZrCoBi} & 270.6  & 77.9   & 58.1  & 142.1  & 71.2  & 1.99      & 183.1   & 0.285    & This work, GGA-PBE    \\
                               & 271.2   & 78.2  & 58.2  & 142.5  & 71.3   &    1.99   & 183.4     & 0.285  & This work, GW
\\ \hline\hline  

\end{tabular}
\end{table*}

For ZrCoSb our predicted Young's modulus is about 2\% higher than the experimental results\cite{sekimoto2005thermoelectric}. The high value of youngs modulus indicates that the covalent bonding component is dominant. The material's stiffness is indicated by a strong covalent bond.

\begin{table}[h]
\caption {The compressional (V$_l$) and shear wave (V$_s$) velocity in m/s, Debye’s temperature $\Theta $ $_{D}$ in K and the shear anisotropy (A) for ZrCoY.} 

\label{tab:table 4}
\begin{tabular}{llllll}
\hline\hline
\textit{\textbf{Comp}}  & \textit{\textbf{V$_l$}} & \textit{\textbf{V$_s$}} & \textit{\textbf{$\Theta $ $_{D}$}} & \textit{\textbf{A}}    & \textit{\textbf{Ref}}                \\
\hline
\multirow{4}{*}{ZrCoSb} & 5589.9 & 3168.6 & 392.8    & 0.72 & This work, GGA-PBE \\
                        & 5620.3 & 5168.5 & 392.9    & 0.71 & This work, GW      \\
                        & 5544.1 & 3379.2 & 392.1    & 0.88 & Ref\cite{joshi2019theoretical}, GGA-PBE       \\
                        & 5488.0    & 3132 .0   & 392.0    &  - & Expt\cite{li2015synthesis}               \\
                        \hline
\multirow{2}{*}{ZrCoBi} & 4801.0 & 2631.2 & 323.7    & 0.63 & This work, GGA-PBE \\
                        & 4804.9 & 2632.6 & 323.9    & 0.60 & This work, GW \\
                        
                        \hline\hline
\end{tabular}
\end{table}

The Debye's temperature ($\Theta $ $_{D}$) , as well as compressional and shear wave velocities, were  computed and the results are in good agreement with the  theoretical and experimental data (where available) as outlined in table \ref{tab:table 4}.  The Debye's temperature is a fundamental parameter that is strongly related to the melting point and specific heats in solids. 

From table \ref{tab:table 4}, the $\Theta $ $_{D}$, in ZrCoSb is greater than in ZrCoBi. $\Theta $ $_{D}$  of ZrCoSb is in excellent agreement with the experimental values in the table \ref{tab:table 4}, however, we report for the first time the $\Theta $ $_{D}$ for ZrCoBi. High Debye's temperatures ($\Theta $ $_{D}$) indicates stiffer crystal orientation, and high melting points are found in such crystals. Since the Debye's temperature ($ \Theta $ $_{D}$) is the temperature required to activate all of a crystal's phonon modes, high $\Theta $ $_{D}$  values suggest that a lot of energy is needed to excite the phonons, making such materials ideal for thermoelectric power generation. Besides, a Debye's temperature of 300 K or higher indicates that the crystal's thermal conductivity is high\cite{heuze2012general}, and since the compounds we are studying have  Debye's temperature of 392.8 K and 323.7 K for ZrCoSb and ZrCoBi, respectively.  We concluded that the thermal conductivity is high.

In addition, we calculated A values of 0.72  and  0.63 for ZrCoSb and ZrCoBi, at GGA-PBE levels, respectively, and  0.71 and 0.60 at GW levels of theory. As indicated in table \ref{tab:table 4}, the calculated values of A are either larger or less than 1 but not equal to 1. It is well known that the value of A is 1 for isotropic crystals. Otherwise, a crystal's shear anisotropy is measured by values less than or greater than 1. As a result, the HH ZrCoY is purely anisotropic.

The ZrCoBi compounds possess lower sound velocities and Young’s modulus compared to  ZrCoSb. Such a low  sound velocity and Young’s modulus originate from the weaker chemical bonding and heavy atomic mass of Bi. The strong relativistic effect of Bi contracts the outer shell and increases its inertness for bonding. Therefore, the lower sound velocity and Young’s modulus  jointly contributes to an intrinsically low lattice thermal conductivity for ZrCoBi.

\subsection{Band structure and Projected Density of States}\label{sec :IIIC}

\begin{figure}[t]
    \centering
    \includegraphics[width=\textwidth]{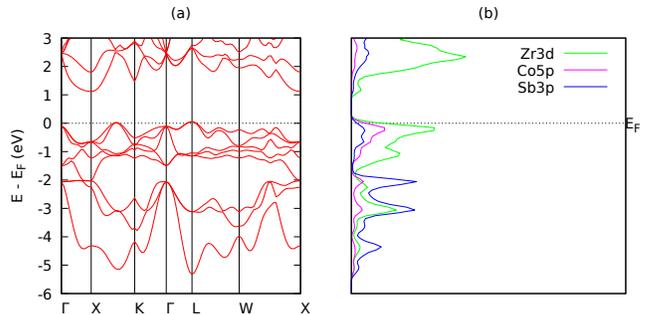}
    \caption{(a) GGA-PBE Calculated Band structures and (b) Projected Density of States (PDOS) for ZrCoSb}
    \label{fig:fig2}
\end{figure}

\begin{figure}[h]
    \centering
    \includegraphics[width=\textwidth]{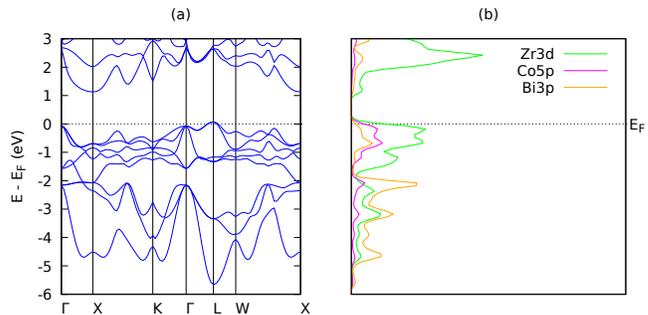}
    \caption{(a) GGA-PBE Calculated Band structures and (b) Projected Density of States (PDOS) for ZrCoBi}
    \label{fig:fig 3}
\end{figure}

The first-principles calculation on the electronic  structures of ZrCoSb (Fig. \ref{fig:fig2}) and ZrCoBi (Fig. \ref{fig:fig 3}) was employed to evaluate their electronic properties. From the electronic band structures in Fig. \ref{fig:fig2}a and Fig. \ref{fig:fig 3}a , the valence band maxima (VBM) locates at $\Gamma$ point, while the conduction band minima (CBm) is at X  point in both compounds. The compounds are indirect $\Gamma$ - X bandgap semiconductors  with a high reflectivity in the infrared portion of the photon energy spectrum, according to electronic band structure analysis and the magnitude of the bandgaps. For indirect bandgap semiconductors, the conductivity is enhanced by phonon vibrations, a detailed discussion on the phonon vibrational properties of these compounds will be discussed later in Sec \ref{sec :IIID}. However, the electronically non-metallic behavior of the ZrCoY compounds has been revealed by a bandgap of slightly above 1 eV. Table \ref{tab:table 5} gives a summary of the calculated bandgaps at GGA-PBE and GW approximations, from the table \ref{tab:table 5}, the GW approximation give a better prediction of the gaps.

\begin{table}[h]
\begin{tabular}{llll} \hline \hline
\multirow{3}{*}{Comp} & \multicolumn{3}{l}{Band gap (eV)}                             \\
                           & \multicolumn{2}{l}{This work} & \multirow{2}{*}{Expt} \\
                           & GGA & GW   &             \\ 
                           \hline  
ZrCoSb                     & 1.05    &     1.44  &      1.45\cite{harrington2017valence}      \\
ZrCoBi                     &  1.03       &  1.35     &    -  \\  
\hline \hline 
\caption{Calculated bandgaps using GGA-PBE in comparison to GW Approximation gaps }
\label{tab:table 5}
\end{tabular}
\end{table}

The projected densities of states in Fig.\ref{fig:fig2}b and Fig.\ref{fig:fig 3}b  demonstrate that Zr dominates both the valence  and conduction bands of the alloys, however, for ZrCoBi, the valence state is dominated by Bi at the energy ranges of -6 to -2 eV. There is also a sharp increase in the density of states near the valence band maximum for the ZrCoBi, indicating high thermal power\cite{jiang2022first}.

\subsection {Lattice dynamics : Phonon Dispersion}\label{sec :IIID}

\begin{figure*}[ht]
     \centering
     \includegraphics[width=0.9\textwidth]{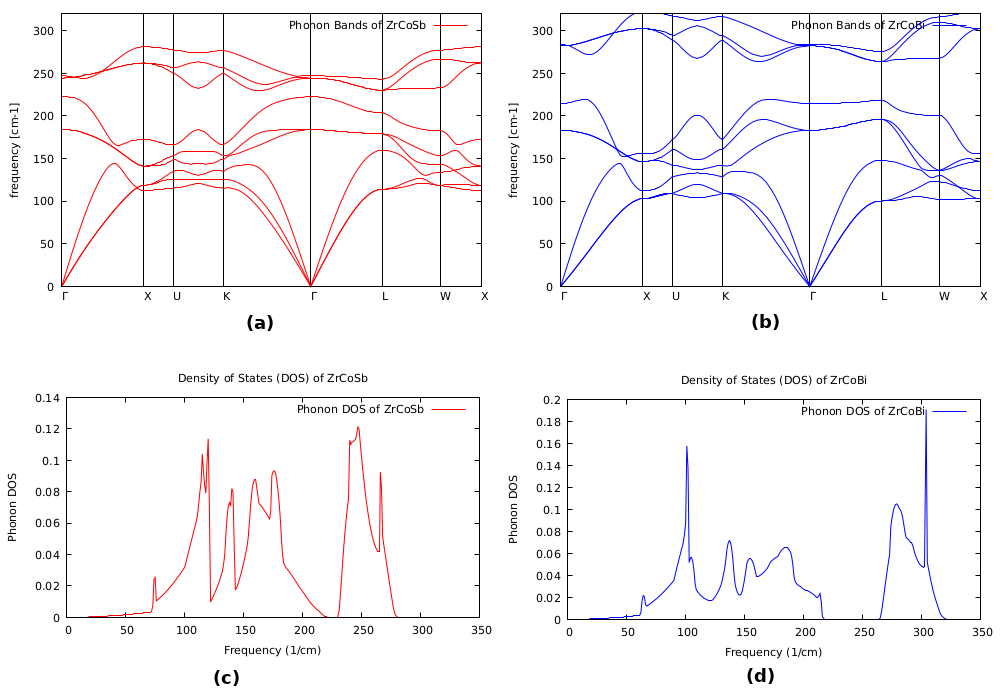}
     \caption{Phonon dispersion Curves for \textbf{(a)}  ZrCoSb, \textbf{(b)} ZrCoBi and Density of states for  \textbf{(c)} ZrCoSb, and \textbf{(d)} ZrCoBi}
     \label{fig:Fig 5}
 \end{figure*}

Phonon frequencies are created when atoms in a crystal are displaced from their rest positions, causing pressures to increase\cite{togo2015first}. Though crystals are expected to be fixed, temperature increase can cause the atoms to vibrate about their mean position. These vibrations influence the possible stability of the alloys, especially at high temperature. The study of the lattice vibration also explains how solids absorb energy. The calculated phonon dispersion curves and density of states for ZrCOSb and ZrCoBi are shown in Fig. \ref{fig:Fig 5}(a)-(d).

We considered the unit cell of the alloys containing three atoms in this investigation. To promote convergence, we built a 2 × 2 × 2 supercell for each of the compounds. The computations occurred in the first Brillouin zone, representing the primitive cell in the reciprocal lattice. The zones sampled in this work covered $\Gamma$ - X - U - K - $\Gamma$- L - W - X.  If the potential energy of a system is continually increasing for any combination of atom displacements, it is considered dynamically stable at equilibrium. Consequently, phonons should have non-negative and real frequencies for stability\cite{togo2015first}. Negative frequencies indicate that the potential energy of the system is decreasing, and so the system is unstable. The two alloys proved to be dynamically stable; this is evident in the fact that from Fig.\ref{fig:Fig 5}(a-d), there are no negative frequencies or dispersions.
Considering that there are three atoms, using the relation \textit{3n} dispersion branches, where \textit{n} is the number of atoms, we observe nine modes in the phonon dispersion spectra in Figs.\ref{fig:Fig 5} (a-b); there are three acoustic modes (3) and six optical modes (3n-3). The optical modes can interact with light. The optical mode for ZrCoBi is between 250–330 cm $ ^{-1}$; this puts the wavelength of the alloy between 30303 nm and 40000 nm while the frequency range for the optical mode in ZrCoSb is between 230-300 cm $^{-1}$. Both the alloys are in the far-infrared region, and a much higher frequency will be required to bring it to the visible region, which is most desirable in solar systems. The maximum frequency for ZrCoBi is slightly above 300 cm $^{-1}$, while ZrCoSb has its maximum frequency slightly lower than 300 cm $^{-1}$, this is attributed to Bi being heavier than Sb. 
A clear gap appears between the acoustic and optical mode for both alloys, and another gap between the two optical modes for the ZrCoSb alloy, these data are required when researching thermal and electrical conductivity as well as determining the thermomelectric properties of materials\cite{giustino2014materials}.
\\
\section{Conclusion}\label{sec :IV}
We report the first-principles results on the structural, electronic, mechanical, and lattice dynamical properties of the HH alloys, ZrCoSb and ZrCoBi.  These properties are calculated within the DFT framework using the GGA-PBE as well as by using GW approximation. The values of elastic constants and mechanical properties obtained from these two approaches are close to one another and also to the available experimental data. However, for band gaps obtained at the GW levels, ZrCoY exhibited  values closer to the experimental data compared to GGA-PBE calculated values. Our multimethod approach clearly shows that
the most elaborate method, i.e.,the GW approach, is required to assess not only the electronic properties, but also the mechanical  properties of the half-Heusler family. The computed values of bulk modulus, shear modulus and Young’s modulus confirmed bond strength, hardness and stiffness of ZrCoY, respectively.  The Bulk modulus values calculated from Thermo-PW code are very close to the values calculated from Murnaghan’s equation of state. For ZiCoBi, we have presented the elastic constants and Debye's temperature for the first time. The two HH alloys have a relatively high Debye's temperature an indication that their thermal conductivity is high. It is deduced from the research that the compounds are mechanically and dynamically stable. The HH alloys ZrCoY (Y=Bi, Sb) are anisotropic and ductile. The compound's properties indicates that they are  good thermal conductors suggesting that an in-depth study of transport properties of ZrCoY using GW approximation is necessary.
 
\section{Acknowledgments}
This research was supported by the Partnership for skills in Applied Sciences, Engineering and
Technology (PASET) which awarded a Regional Scholarship and Innovation Fund (RSIF). The support from the International Science Program (KEN02), through the Department of Physics, University of Nairobi, is acknowledged. The Centre of High-Performance Computing (CHPC) through the project MATS862, Rosebank Cape Town Republic of South Africa is appreciated for providing access to the HPC cluster facility used in this research. Much appreciation to computational modelling team at Egerton University, under the mentorship of Dr. Winfred Mulwa, for the insightful discussions offered.

\section{Data availability}
All the data in this study are available and can be made available on request with the corresponding author.

\section{conflicts of interest}
There is no conflict of interest.

\bibliography{lyn}
\end{document}